\documentclass[aps,prd,showpacs,twocolumn,groupedaddress]{revtex4}
\usepackage{graphicx}

\begin{document}

\title{Lattice QCD study of the heavy-heavy-light quark potential}

\author{Arata~Yamamoto}
\affiliation{Faculty of Science, Kyoto University, Kitashirakawa, Sakyo, Kyoto 606-8502, Japan}

\author{Hideo~Suganuma}
\affiliation{Faculty of Science, Kyoto University, Kitashirakawa, Sakyo, Kyoto 606-8502, Japan}

\author{Hideaki~Iida}
\affiliation{Yukawa Institute for Theoretical Physics, Kyoto University, Kitashirakawa, Sakyo, Kyoto 606-8502, Japan}

\date{\today}

\begin{abstract}
We study the heavy-heavy-light quark ($QQq$) potential in SU(3) quenched lattice QCD, and discuss one of the roles of the finite-mass valence quark in the inter-quark potential.
Monte Carlo simulations are performed with the standard gauge action on the $16^4$ lattice at $\beta =6.0$ and the $O(a)$-improved Wilson fermion action at four hopping parameters.
For statistical improvement, the gauge configuration is fixed with the Coulomb gauge.
We calculate the potential energy of $QQq$ systems as a function of the inter-heavy-quark distance $R$ in the range of $R \le$ 0.8 fm.
The $QQq$ potential is well described with a Coulomb plus linear potential, and the effective string tension between the two heavy quarks is significantly smaller than the string tension $\sigma \simeq 0.89$ GeV/fm. 
It would generally hold that the effect of the finite-mass valence quark reduces the inter-two-quark confinement force in baryons.
\end{abstract}


\maketitle

\section{INTRODUCTION}
In hadron physics, the inter-quark interaction is one of the fundamental and essential properties.
In particular, the quark confinement is not only an important property of hadrons but also an important problem of the modern physics.
The non-Abelian and strong coupling nature of QCD makes it difficult to treat the inter-quark interaction and many other nonperturbative phenomena analytically.
Lattice QCD is the first-principle calculation based on the QCD Lagrangian, and is one of the most useful approaches for such nonperturbative phenomena \cite{Cr81,Ro92}.
 
The quark confinement in hadrons is  well described by the picture of the gluonic ``flux tube" or ``string" \cite{Na74}.
This means that the confinement potential is a linear function of the flux-tube length.
For example, the quark-antiquark ($Q\bar Q$) potential is written as the one-gluon-exchange Coulomb potential plus the linear confinement potential,
\begin{eqnarray}
V_{Q\bar Q}(R) = \sigma_{Q\bar Q}R -\frac{A_{Q\bar Q}}{R}+C_{Q\bar Q},
\label{VQbarQ}
\end{eqnarray}
which is called the Cornell potential \cite{Ei78}.
Here $R$ is the distance between the quark and the antiquark, and it is equal to the gluonic flux-tube length of the $Q\bar Q$ system.
Lattice QCD also reproduces this functional form of the $Q\bar Q$ potential \cite{Cr7980,Ba92}.

In addition, the lattice QCD calculations reveal that this picture holds in three-quark ($3Q$) systems and multi-quark systems.
The $3Q$ potential is obtained from quenched lattice QCD as
\begin{eqnarray}
V_{3Q}(\vec{r}_1,\vec{r}_2,\vec{r}_3)=\sigma _{3Q}L_{\rm min}-\sum _{i< j}\frac{A_{3Q}}{|\vec{r}_i-\vec{r}_j|}+C_{3Q},
\label{V3Q}
\end{eqnarray}
where $(\vec{r}_1,\vec{r}_2,\vec{r}_3)$ are the coordinates of three quarks and $L_{\rm min}$ is the flux-tube length \cite{Ta0102}.
The $3Q$ flux tube has the geometry minimally connecting the three quarks, and forms the Y-type structure \cite{Ta0102,Ic03,Br95,Co04}.
The similar potentials are obtained in multi-quark systems \cite{Ok05}.
In the flux-tube picture of $3Q$ and multi-quark systems, the confining force is a many-body force, reflecting the complicated gluonic dynamics based on the SU(3) gauge symmetry. 
The strength of the confinement by these flux tubes, i.e., the string tension, is the universal value in hadrons, about 0.89 GeV/fm.

These inter-quark interactions are mainly constructed from the gluon dynamics.
In particular, as most of previous lattice works, the static and quenched calculation strictly gives only the gluonic potential.
However, since not only gluons but also quarks exist in hadrons, the realistic inter-quark potential would include quark effects.
The quark effects can be categorized into two types.

One is the unquenched effect, or the sea quark effect.
As already known, the sea quark causes ``string breaking" and flattens the potential slope in long range.
This phenomenon is convinced also in lattice QCD \cite{Ba05}.

The other is the finite-mass valence quark effect.
An example of this effect is the relativistic correction to the Coulomb interaction, i.e., the Fermi-Breit interaction \cite{Ru75}.
As well as the Coulomb potential, we can also expect the finite-mass quark effects on the confinement potential.
In 3Q or multi-quark systems, there would exist nontrivial effects reflecting the characteristic flux-tube structure.
In this paper, we investigate this ``motional" effect of finite-mass valence quarks.

\begin{figure*}[t]
\begin{center}
\includegraphics[scale=0.5]{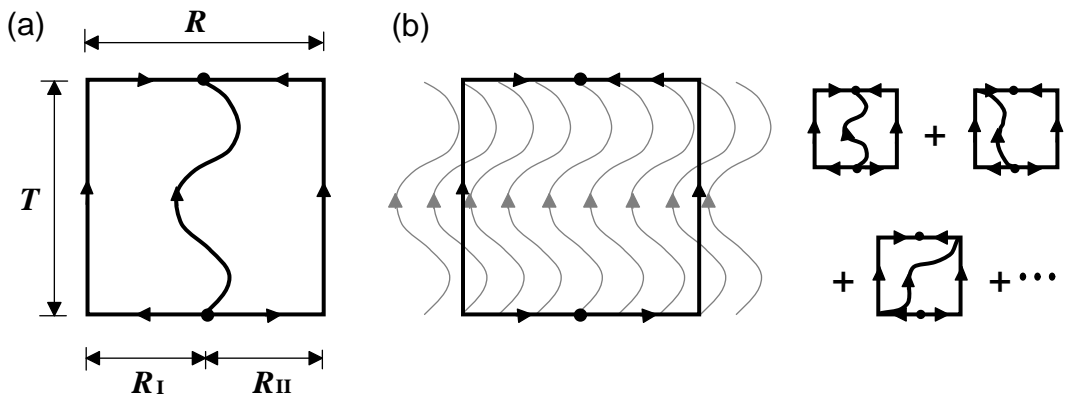}
\caption{\label{fig1}
(a)The $QQq$ Wilson loop.
The wavy line represents the light-quark propagator and the straight line the heavy-quark trajectory.
(b)The ``wall-to-wall $QQq$ Wilson loop."
The gray wavy lines represent the wall-to-wall quark propagator, which is the average of all propagators at a fixed time separation.
}
\end{center}
\end{figure*}

To investigate such quark motional effects, we define the heavy-heavy-light quark ($QQq$) potential \cite{YaL}.
Consider the $QQq$ system which is constructed with two static quarks and one finite-mass quark.
The $QQq$ potential $V_{QQq}(R)$ is defined as the energy of the $QQq$ system in terms of the inter-heavy-quark distance $R$.
It is the inter-two-quark potential in baryons which effectively includes the light-quark effects.
We calculate the $QQq$ potential in lattice QCD, and study the nontrivial light-quark effect by comparing it with the static $Q\bar Q$ or $3Q$ potential.

As an example of experimental $QQq$ baryons, the doubly charmed baryon is recently observed by the SELEX Collaboration at Fermilab.
In 2002, they observed the doubly charmed baryon $\Xi _{cc}^+(dcc)$ through a decay process $\Xi _{cc}^+ \to \Lambda _c^+ K^- \pi ^+$ \cite{Ma02}.
They also confirmed another decay process $\Xi _{cc}^+ \to p D^+ K^- $ \cite{Oc05}.
However, there is also the negative result, for example, by the BABAR Collaboration \cite{Au06}.
The experimental discovery of the doubly charmed baryons is still under debate, and more experimental information is desired.
The doubly charmed baryons are also theoretically investigated in lattice QCD \cite{Le0102} and other approaches \cite{Ru75,Vi04,Br05}.

In this paper, we calculate the $QQq$ potential in SU(3) lattice QCD, and investigate the finite-mass valence quark effect in the inter-quark potential in baryons.
In Sec.~II, we formulate the potential calculation using the $QQq$ Wilson loop.
In Sec.~III, we introduce the Coulomb gauge fixing for the statistical error reduction, and consider its effect on the inter-quark potential.
In Sec.~IV, we show the lattice QCD data and the resulting $QQq$ potential, and discuss the finite-mass valence quark effect on the $QQq$ potential.
Section V is devoted to the summary.

\section{LATTICE QCD FORMALISM}
\subsection{$QQq$ Wilson loop}

As the basic potential calculations in lattice QCD, we define the $QQq$ Wilson loop for the $QQq$ potential calculation.
As shown in Fig.~\ref{fig1}(a), the $QQq$ Wilson loop is constructed from two ``staples" of static-quark trajectories and one light-quark propagator.
The two staples $U^{\rm I}$ and $U^{\rm II}$ are the path-ordered product of link variables, as the static inter-quark potential calculation.
The major difference from the usual potential calculation is that the $QQq$ Wilson loop includes the quark propagator $K^{-1}$, which represents the light quark moving around. 
They are written as
\begin{eqnarray}
\label{WQQq}
W_{QQq} (R,T)&\equiv& \frac{1}{3!}\epsilon_{abc}\epsilon_{def} U^{\rm I}_{ad} U^{\rm II}_{be} K_{cf}^{-1},\\ 
U^{k} &=& Pe^{ig\int_{\Gamma _k} dx^\mu A_\mu} \qquad(k={\rm I,II}),\\
K_{ab}^{-1} &=& \int D\bar{q} Dq \, q_a \bar{q}_b e^{-\bar{q} K q},
\end{eqnarray}
where the subscripts $a,b,...,f$ are color indices.
When $T$ is large enough, the expectation value of the $QQq$ Wilson loop should depend only on the spatial size $R$ and the temporal size $T$, not on the position of the source and sink junction points ($R_{\rm I}$ and $R_{\rm II}$ in Fig.~\ref{fig1}(a)).
The resulting $QQq$ potential is written with one parameter $R$, which is the distance between the two heavy quarks.

The $QQq$ potential is obtained from the expectation value of the $QQq$ Wilson loop as
\begin{eqnarray}
V_{QQq} (R)= - \lim _{T\rightarrow \infty} \frac{1}{T}\ln \langle W_{QQq}(R,T) \rangle.
\label{lnW}
\end{eqnarray}
The symbol $\langle \quad \rangle$ means the expectation value integrated over the gauge field.
In practical analysis, the expectation value is fitted with a single exponential form $Ce^{-V_{QQq}T}$ in large but finite range of $T$.
To estimate the suitable fit range of $T$, the effective mass is defined as
\begin{eqnarray}
v(R,T)\equiv \ln \frac{\langle W_{QQq}(R,T) \rangle}{\langle W_{QQq}(R,T+1)\rangle}.
\end{eqnarray}
If the state is dominated by a single component of the potentials, the effective mass is independent of $T$.
The plateau region of the effective mass is the indication to determine the fit range of $T$.
We calculate the expectation value in lattice QCD for several values of $R$, and explore a suitable functional form of $V_{QQq}(R)$.

In Eq.~(\ref{WQQq}) and Fig.~\ref{fig1}(a), the $QQq$ Wilson loop is defined as a single gauge-invariant loop.
For the reduction of the statistical error, we actually use the ``wall-to-wall $QQq$ Wilson loop".
We define the wall-to-wall quark propagator as
\begin{eqnarray}
K^{-1}_{\rm wall}(T)\equiv \frac{1}{V^2}\sum_{n_{\rm src}}\sum_{n_{\rm sink}} K^{-1}(n_{\rm src},n_{\rm sink},T),
\end{eqnarray}
where $n_{\rm src}$ and $n_{\rm sink}$ are the spatial sites of the source and sink respectively, and $V$ is the number of the spatial lattice sites.
This wall-to-wall propagator is the averaged propagator from the whole space at one time to that at another time.
(The ``wall" means the average over all spatial sites.)
The wall-to-wall $QQq$ Wilson loop is constructed by replacing the single quark propagator in the $QQq$ Wilson loop with this wall-to-wall quark propagator.
The schematic figure is depicted in Fig.~\ref{fig1}(b).
Because such a propagator is independent of the spatial position, we can easily sum up the parallel translated wall-to-wall $QQq$ Wilson loops in the whole space.
This summing up drastically suppresses the statistical error owing to the large statistics.
Without gauge fixing, the disconnected components in the wall-to-wall $QQq$ Wilson loops automatically vanish by Elitzur's theorem and only the gauge-invariant components remain, and therefore the resulting $QQq$ potential is gauge invariant.

\subsection{Simulation details}

\begin{table}[t]
\caption{\label{tab1}
Simulation parameters. The list shows $\beta =2N_c/g^2$ and the corresponding lattice spacing $a$, the sweep numbers ($N_{\rm therm},N_{\rm sep}$) of the thermalization and separation for updating the gauge fields, the smearing parameters ($\alpha,N_{\rm smr}$), and the clover coefficient $c$.
}
\begin{ruledtabular}
\begin{tabular}{cccccccc}
$\beta$ & $a$ [fm] & lattice size & $N_{\rm therm}$ &
$N_{\rm sep}$ & $\alpha$ & $N_{\rm smr}$ & $c$ \\
\hline
6.0 & 0.10 & $16^4$ & 10000 & 500 & 2.3 & 40 & 1.479\\
\end{tabular}
\end{ruledtabular}

\caption{\label{tab2}
The correspondence between $\kappa$ and the meson masses.
The list shows the used gauge configuration number $N_{\rm conf}$, the pion mass $m_{\pi}$, the $\rho$ meson mass $m_{\rho}$, and the approximate constituent quark mass $M_q\simeq m_{\rho}/2$.
The statistical error is estimated with the jackknife method.
}
\begin{ruledtabular}
\begin{tabular}{ccccc}
$\kappa$ & $N_{\rm conf}$ & $m_{\pi}$ & $m_{\rho}$ & $M_q$ \\
\hline
0.1200 & 1000 & 1.446(1) & 1.472(2) &  1.5 GeV\\
0.1300 &  300 & 0.900(2) & 0.949(1) &  1 GeV\\
0.1340 &  300 & 0.643(1)  & 0.716(1) &  700 MeV\\
0.1380 & 1000 & 0.304(1) & 0.467(2)&  500 MeV\\
\end{tabular}
\end{ruledtabular}
\end{table}

We perform SU(3) lattice QCD calculation at the quenched level.
Simulation parameters are summarized in Table \ref{tab1}.
The gauge action is the standard isotropic plaquette action and $\beta=6.0$.
The corresponding lattice spacing $a\simeq 0.10$ fm is determined so as to reproduce the string tension $\sigma_{Q\bar Q}$ of the $Q\bar Q$ potential to be 0.89 GeV/fm.
We use this lattice unit for most part of the paper.
Lattice volume is $16^4$, and periodic boundary conditions are imposed on the space-time boundaries. 
For generating the gauge configurations, we use the pseudo-heat-bath algorithm and take 10000 sweeps for the thermalization, and 500 sweeps for the separation of each configuration.

For the ground-state component enhancement, we apply the smearing method \cite{Ta0102,Bo92} to the spatial link variables of the $QQq$ Wilson loop. 
In the smearing method for SU(3) link variables, we iteratively replace the link variable $U_i(n)$ $(i=1,2,3)$ with $\tilde{U}_i(n)(\in $ SU(3)) which maximizes
\begin{eqnarray}
&{\rm ReTr} \Bigl[ \tilde{U}_i^{\dagger}(n) \bigl\{ \alpha U_i(n) + \sum _{j\neq i} \bigl( U_j(n)U_i(n+\hat{j})U_j^\dagger (n+\hat{i})\nonumber\\
&+U_j^\dagger (n-\hat{j})U_i(n-\hat{j})U_j(n+\hat{i}-\hat{j}) \bigr) \bigr\} \Bigr]. 
\end{eqnarray}
Such a replacement does not change the gauge transformation property of the link variable, and this smearing method is a gauge-invariant manner.
Physically, the gauge-invariant smearing method changes a stringy link to a spatially-extended flux tube.
It does not change the physical content such as the potential, and enhances the ground-state component so that the statistical error is suppressed.
The method has two parameters, a real parameter $\alpha$ and the iteration number $N_{\rm smr}$ of the replacements, and our choice of $\alpha$ and $N_{\rm smr}$ is based on the static $3Q$ case \cite{Ta0102}.

As for the fermion action for the light-quark propagator, we adopt the $O(a)$-improved Wilson fermion action, i.e., the clover fermion action \cite{El97} 
\begin{eqnarray}
S_{\rm quark}&=&\sum _{n,m} \bar{q}_n K_{nm} q _m,\\
K_{nm}&=&\delta _{n,m} -\kappa \sum _\mu \{ (1-\gamma _\mu)U_\mu (n)\delta _{n+\hat{\mu},m}\nonumber \\
&&+(1+\gamma _\mu)U_\mu^\dagger (n-\hat{\mu})\delta _{n-\hat{\mu},m} \}\nonumber \\
&&-\kappa c\sum _{\mu <\nu} \sigma _{\mu \nu}F_{\mu \nu} \delta _{n,m},
\end{eqnarray}
where $n$ and $m$ are the space-time site indices, and other indices are omitted.
The clover coefficient $c$ in this action is given from the mean field value $u_0$ of the link variable for the tadpole improvement.
We determine $c$ and $u_0$ from the plaquette value $P_{\mu\nu}(n)$ as
\begin{eqnarray}
c=\frac{1}{u_0^3}, && u_0=\Bigl\langle \frac{1}{3}{\rm ReTr} P_{\mu\nu}(n) \Bigr\rangle ^{\frac{1}{4}}.
\end{eqnarray}
Here $\langle \quad \rangle$ means the average of all plaquettes in the gauge configurations as the ensemble average.
The measured mean field value $u_0$ is 0.87779(2) in our gauge configurations.
To investigate the light-quark-mass dependence, we take different four light-quark hopping parameters, $\kappa =0.1200$, 0.1300, 0.1340, and 0.1380.
In Table \ref{tab2}, we list the results of the meson correlator calculations using these hopping parameters.
The constituent quark mass $M_q$ is roughly estimated with the half of the $\rho$ meson mass $m_\rho$.
Our calculations cover the mass region of 0.5 GeV $\le M_q \le 1.5$ GeV.

The Monte Carlo simulations are performed on NEC SX-8R at Osaka University.

\section{COULOMB GAUGE FIXING}
\subsection{Statistical error reduction}

In principle, the $QQq$ potential can be calculated with the formalism which is mentioned in the Section II.
However, since the statistical error is severely large, some improvements are necessary for statistical error reduction.
We have tried several methods for the error reduction, and adopted the Coulomb gauge fixing procedure.
In this procedure, the light-quark propagator and the heavy-quark trajectories in the wall-to-wall $QQq$ Wilson loop is calculated after the Coulomb gauge fixing. 
The Coulomb gauge fixing procedure surprisingly suppresses the statistical error in this study.

Since the wall-to-wall propagator is a spatial average of all propagators, the wall-to-wall $QQq$ Wilson loop includes disconnected loops, where the two staples and the light-quark propagator are not connected at the source and sink junction points.
The expectation values of such disconnected loops vanish in gauge-invariant formalism, but do not vanish with the Coulomb gauge. 
Such contributions are gauge-variant artifacts by the nonlocal nature of the Coulomb gauge, and can give the different result from the gauge-invariant potential.
However, empirically, the gauge-variant components at the Coulomb gauge rapidly decrease and do not affect the resulting potential \cite{Gr03}.

In this section, we show several numerical results of $Q\bar Q$ and $3Q$ potentials with the Coulomb gauge, and check that its influence is small enough for inter-quark potential calculations.
Other statistical improvement methods which we have tried are added to Appendix A.

\subsection{$Q\bar Q$ potential with the Coulomb gauge}
First, we consider the $Q\bar Q$ potential with the Coulomb gauge \cite{Gr03}.
We set the correlator of two Wilson lines
\begin{eqnarray}
W_{Q\bar Q}^{\rm Coul}(R,T)\equiv \frac{1}{3}{\rm Tr}\{ L(\vec{r}_1,T) L^\dagger(\vec{r}_2,T) \},
\end{eqnarray}
where $L(\vec{r},T)$ is a Wilson line with temporal length $T$ and $R=|\vec{r}_1-\vec{r}_2|$.
The expectation value of this correlator is nonzero with the Coulomb gauge.
The $Q\bar Q$ potential $V_{Q\bar Q}^{\rm Coul}$ with the Coulomb gauge is extracted by fitting with $\langle W_{Q\bar Q}^{\rm Coul}\rangle =C{\rm exp}(-V_{Q\bar Q}^{\rm Coul}T)$.
The best-fit result of the on-axis data with $V_{Q\bar Q}^{\rm Coul}(R)=\sigma_{Q\bar Q}^{\rm Coul}R-A_{Q\bar Q}^{\rm Coul}/R+C_{Q\bar Q}^{\rm Coul}$ is listed in Table \ref{tab3}.
Although the potential seems to include some gauge-dependent contributions in small $T$, when the fit range of $T$ is large enough, gauge artifacts dump and the potential approaches the physical $Q\bar Q$ potential.
Compared to the gauge-invariant result in SU(3) lattice QCD with $\beta =6.0$ \cite{Ba92}, the string tension $\sigma_{Q\bar Q}^{\rm Coul}$ is close to or slightly higher than the physical value $\sigma_{Q\bar Q}= 0.0534(18)$, and the Coulomb coefficient $A_{Q\bar Q}^{\rm Coul}$ is almost the same as the physical value $A_{Q\bar Q}= 0.267(6)$.

\begin{table}[b]
\caption{\label{tab3}
The best fit parameters of the on-axis $Q\bar Q$ potential with the Coulomb gauge.
The list has different fit ranges of $T$ for fitting with $\langle W_{Q\bar Q}^{\rm Coul}\rangle =C{\rm exp}(-V_{Q\bar Q}^{\rm Coul}T)$.
$N_{\rm dof}$ is the degree of freedom.
}
\begin{ruledtabular}
\begin{tabular}{ccccc}
fit range & $\sigma_{Q\bar Q}^{\rm Coul}$ & $A_{Q\bar Q}^{\rm Coul}$ & $C_{Q\bar Q}^{\rm Coul}$ & $\chi ^2/N_{\rm dof}$\\
\hline
$T$=[1,8] & 0.062(3) & 0.258(10) & 0.609(1) & 3.13\\
$T$=[2,8] & 0.060(2) & 0.254(7) & 0.606(9) & 2.86\\
$T$=[3,8] & 0.058(2) & 0.256(7) & 0.609(2) & 2.69\\
$T$=[4,8] & 0.059(1) & 0.253(5) & 0.605(3) & 0.41\\
\end{tabular}
\end{ruledtabular}

\caption{\label{tab4}
The $3Q$ potential with the Coulomb gauge extracted from $W_{3Q\rm B}^{\rm Coul}$.
The best-fit parameters are of the function (\ref{V3QC}) with different fit ranges of $T$.
}
\begin{ruledtabular}
\begin{tabular}{ccccc}
fit range & $\sigma_{3Q}^{\rm Coul}$ & $A_{3Q}^{\rm Coul}$ & $C_{3Q}^{\rm Coul}$ & $\chi^2/N_{\rm dof}$ \\
\hline
$T=[1,8]$ & 0.0498(3) & 0.138(1) & 0.957(3) & 6.14\\
$T=[2,8]$ & 0.0474(3) & 0.138(1) & 0.959(3) & 8.34\\
$T=[3,8]$ & 0.0466(7) & 0.141(2) & 0.968(5) & 3.91\\
$T=[4,8]$ & 0.0482(15) & 0.136(3) & 0.950(11) & 3.95\\
\end{tabular}
\end{ruledtabular}
\end{table}

\subsection{$3Q$ potential with the Coulomb gauge}

\begin{figure}[t]
\begin{center}
\includegraphics[scale=0.4]{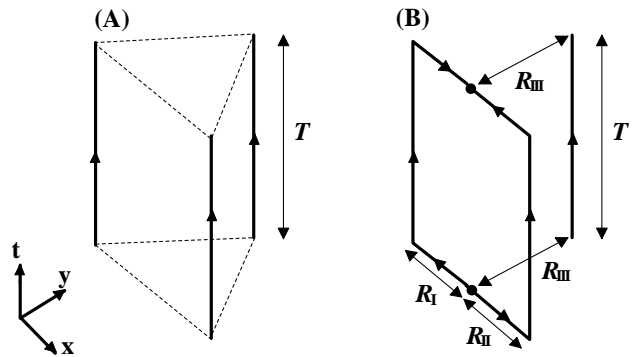}
\caption{\label{fig2}
The correctors $W_{3Q\rm A}^{\rm Coul}$ (left) and $W_{3Q\rm B}^{\rm Coul}$ (right) for the $3Q$ potential with the Coulomb gauge.
$R_{\rm I}$ and $R_{\rm II}$ are the same in Fig.~\ref{fig1}(a), and $R_{\rm III}$ is the distance between the two staples and the Wilson line. 
}
\end{center}
\end{figure}

\begin{figure*}[t]
\begin{center}
\includegraphics[scale=1]{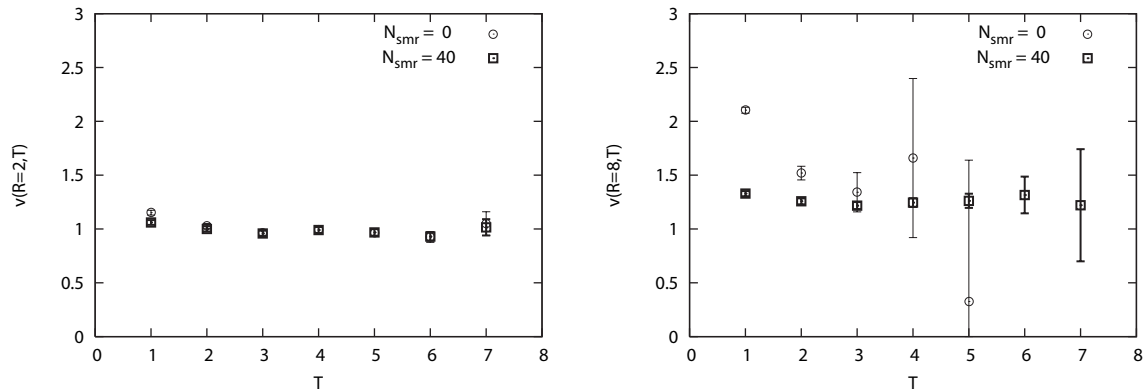}
\caption{\label{fig3}
The effective mass of the wall-to-wall $QQq$ Wilson loop with the Coulomb gauge and the efficiency of the smearing method.
The lightest quark case, $\kappa =0.1380$, is shown.
The left graph is the small loop case with $R=2$ and $(R_{\rm I},R_{\rm II})=(1,1)$ and the right is the large loop case with $R=8$ and $(R_{\rm I},R_{\rm II})=(4,4)$.
In each graph, two kinds of data are the result without the smearing method (circle) and with the smearing method of $(\alpha,N_{\rm smr})=(2.3,40)$ (square).
Thus, the smearing method is found to be effective, especially for large $R$.
}
\end{center}
\end{figure*}

Next, we turn to the $3Q$ case, which is more relevant for our $QQq$ potential calculation.
As for the $3Q$ potential with the Coulomb gauge, we consider two types of correlators.
One is the simple correlator of three Wilson lines
\begin{eqnarray}
W_{3Q\rm A}^{\rm Coul} \equiv \frac{1}{3!}\epsilon_{abc}\epsilon_{def} L_{ad}(\vec{r}_1,T) L_{be}(\vec{r}_2,T) L_{cf}(\vec{r}_3,T), 
\end{eqnarray}
and the other is the correlator of one Wilson line and two ``staples" $U^{\rm I}$ and $U^{\rm II}$
\begin{eqnarray}
W_{3Q\rm B}^{\rm Coul} \equiv \frac{1}{3!}\epsilon_{abc}\epsilon_{def} U^{\rm I}_{ad} U^{\rm II}_{be} L_{cf}(\vec{r}_3,T), 
\end{eqnarray}
which is a closer geometry to the $QQq$ Wilson loop.
The schematic figures are shown in Fig.~\ref{fig2}.
There is not an essential difference between the results from these correlators, but $W_{3Q\rm B}^{\rm Coul}$ has an advantage in accuracy since the smearing method is available.

The numerical results extracted from $W_{3Q\rm B}^{\rm Coul}$ is shown in Table \ref{tab4}.
The total number of geometries of $W_{3Q\rm B}^{\rm Coul}$ is 96: $0\le R_{\rm I}\le 4$, $0\le R_{\rm II}\le 4$, and $1\le R_{\rm III}\le 4$, except for $R_{\rm I}=R_{\rm II}=0$.
The $3Q$ potential with the Coulomb gauge can be written with the same form as the physical $3Q$ potential,
\begin{eqnarray}
V_{3Q}^{\rm Coul}=\sigma _{3Q}^{\rm Coul}L_{\rm min}-\sum _{i< j}\frac{A_{3Q}^{\rm Coul}}{|\vec{r}_i-\vec{r}_j|}+C_{3Q}^{\rm Coul}.
\label{V3QC}
\end{eqnarray}
The $3Q$ string tension $\sigma_{3Q}^{\rm Coul}$ and the Coulomb coefficient $A_{3Q}^{\rm Coul}$ are almost the same as the physical values $\sigma_{3Q}= 0.0460(4)$ and $A_{3Q}= 0.1366(11)$ \cite{Ta0102}.
Considering artifacts from the geometrical asymmetry, we have also tried a more general fit function
\begin{eqnarray}
V_{3Q}^{\rm Coul}=\sigma _{3Q}^{\rm Coul}L_{\rm min}-\sum _{i< j}\frac{A_{ij}^{\rm Coul}}{|\vec{r}_i-\vec{r}_j|}+C_{3Q}^{\rm Coul},
\end{eqnarray}
where there are five fit parameters.
In this case, we have found that $A_{12}^{\rm Coul}\simeq A_{13}^{\rm Coul}\simeq A_{23}^{\rm Coul}$ and the result is unchanged from the fit function (\ref{V3QC}).

From above results, the $Q\bar Q$ and $3Q$ potentials with the Coulomb gauge are found to approach the physical potentials if the fit range of $T$ is large enough.
This would hold both in the long-range physics, such as the string tension, and in the short-range physics, such as the one-gluon-exchange Coulomb potential.
Therefore, we can expect that the physical $QQq$ potential is approximately obtained from the wall-to-wall $QQq$ Wilson loop with the Coulomb gauge in the whole region.

\section{LATTICE QCD RESULTS}
\subsection{$QQq$ potential}

In Fig.~\ref{fig3}, we plot typical examples of the effective mass of the wall-to-wall $QQq$ Wilson loop with the Coulomb gauge.
We show the smallest-quark-mass case, $\kappa =0.1380$, where the statistical fluctuation is the largest.
In the figure, we compare two kinds of data with and without the gauge-invariant smearing method.
The smearing method enhances the ground state component of $QQq$ potential, especially in the large loop case.
Then the effective mass is almost flat in $T \ge 3$ and thus the ground-state component dominates.

\begin{figure}[t]
\begin{center}
\includegraphics[scale=1.1]{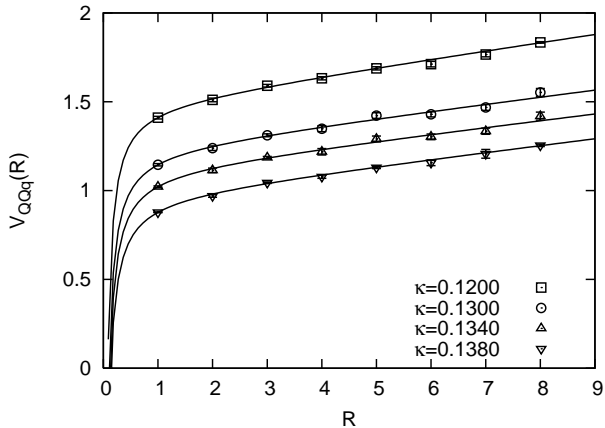}
\caption{\label{fig4}
The lattice QCD data of $QQq$ potential $V_{QQq}$ with the Coulomb gauge.
The results of different four hopping parameters $\kappa$ are shown.
The solid curves are the best-fit functions of Eq.~(\ref{VQQq}).
All the scales are measured in lattice unit.
}
\end{center}
\end{figure}

In the region where the effective mass is flat, we fit $\langle W_{QQq} (R,T)\rangle$ with a single exponential form $C{\rm exp}(-V_{QQq}T)$, and obtain the $QQq$ potential $V_{QQq}(R)$.
The resulting values with $\kappa=0.1380$ are listed in Table \ref{tab5}, and $\kappa=0.1200$ in Table \ref{tab6}.
As mentioned above, $V_{QQq}$ is almost independent of the fit range of $T$, and then the ground state component dominates.
It is also confirmed that $V_{QQq}$ does not depend on the position of the junction points $R_{\rm I}$ and $R_{\rm II}$ separately, but depends only on $R$.
Then we can describe the $QQq$ potential as a function of the inter-heavy-quark distance $R$.
The same arguments hold also in other two $\kappa$ cases.

For a functional form of $V_{QQq}(R)$, we consider the Coulomb plus linear potential,
\begin{eqnarray}
V_{QQq}(R)=\sigma _{\rm eff}R-\frac{A_{\rm eff}}{R}+C_{\rm eff},
\label{VQQq}
\end{eqnarray}
as the analogy of the $Q\bar Q$ potential (\ref{VQbarQ}).
The subscript ``eff" means these values effectively including the light-quark effect.
This simple function is surprisingly suitable for $V_{QQq}(R)$, and the best-fit parameters and the resulting potential form are shown in Table \ref{tab7} and  Fig.~\ref{fig4}, respectively.
For comparison, the string tension and the Coulomb coefficient in the static $3Q$ potential (\ref{V3Q}) are
\begin{eqnarray}
\sigma _{3Q} \simeq  0.045, \quad
A_{3Q} \simeq  0.13
\end{eqnarray}
in the lattice unit at $\beta =6.0$ \cite{Ta0102}.
The Coulomb coefficient $A_{\rm eff}$ is almost the same value as $A_{3Q}$, which is consistent with the short-distance behavior of the $3Q$ potential in perturbative QCD.
In contrast, $\sigma _{\rm eff}$ is about 10-20\% reduced compared to $\sigma_{3Q}$ at $\kappa =0.1300$, 1340, and 1380, as
\begin{eqnarray}
\sigma_{\rm eff} < \sigma _{3Q}.
\end{eqnarray}
In the heaviest case, $\kappa =0.1200$, the effective string tension approximately equals to the string tension.
Let us call this parameter $\sigma _{\rm eff}$ the ``effective string tension".
We have found that the effective string tension is smaller than the string tension of the static $3Q$ system.
The effective string tension strongly depends on $\kappa$, or the light quark mass.

\begin{table}[t]
\newcommand{\m}{\hphantom{$-$}}
\newcommand{\cc}[1]{\multicolumn{1}{c}{#1}}
\renewcommand{\tabcolsep}{0.6pc} 
\renewcommand{\arraystretch}{1} 
\caption{\label{tab5}
The lattice QCD results for the $QQq$ potential $V_{QQq}$ with the Coulomb gauge at $\kappa =0.1380$. 
$R$ and $(R_{\rm I},R_{\rm II})$ denote the loop size defined in Fig.~\ref{fig1}.
The results with different fit ranges of $T$ are also shown.
All the values are in lattice unit, and the statistical error is estimated with the jackknife method.
}
\begin{center}
\begin{tabular}{cccccc}
\hline\hline
$R$ & $(R_{\rm I},R_{\rm II})$ & $V_{QQq}$ & $V_{QQq}$\\
&& $T=[4,8]$ & $T=[5,8]$\\
\hline
1 & (0,1) & 0.877(2) & 0.873(2)\\
2 & (0,2) & 0.971(7) & 0.959(9)\\
  & (1,1) & 0.969(8) & 0.958(10)\\
3 & (0,3) & 1.047(4) & 1.045(7)\\
  & (1,2) & 1.045(4) & 1.043(8)\\
4 & (0,4) & 1.083(11) & 1.067(17)\\
  & (1,3) & 1.079(10) & 1.063(16)\\
  & (2,2) & 1.078(10) & 1.063(15)\\
5 & (0,5) & 1.136(6) & 1.122(3)\\
  & (1,4) & 1.131(6) & 1.117(4)\\
  & (2,3) & 1.130(6) & 1.116(5)\\
6 & (0,6) & 1.170(13) & 1.151(24)\\
  & (2,4) & 1.157(16) & 1.136(30)\\
  & (3,3) & 1.157(16) & 1.136(31)\\
7 & (0,7) & 1.219(21) & 1.220(50)\\
  & (3,4) & 1.207(24) & 1.209(60)\\
8 & (0,8) & 1.262(11) & 1.283(21)\\
  & (4,4) & 1.255(6) & 1.271(10)\\
\hline\hline
\end{tabular}
\end{center}
\end{table}

\begin{table}[t]
\newcommand{\m}{\hphantom{$-$}}
\newcommand{\cc}[1]{\multicolumn{1}{c}{#1}}
\renewcommand{\tabcolsep}{0.6pc} 
\renewcommand{\arraystretch}{1} 
\caption{\label{tab6}
The lattice QCD results for the $QQq$ potential $V_{QQq}$ with the Coulomb gauge at $\kappa =0.1200$. 
The notations are the same as Table \ref{tab5}.
}
\begin{center}
\begin{tabular}{cccc}
\hline\hline
$R$ & $(R_{\rm I},R_{\rm II})$ & $V_{QQq}$ & $V_{QQq}$\\
&& $T=[4,8]$ & $T=[5,8]$\\
\hline
1 & (0,1) & 1.410(7) & 1.398(4)\\
2 & (0,2) & 1.512(10)  & 1.492(6)\\
  & (1,1) & 1.510(10)  & 1.491(7)\\
3 & (0,3) & 1.593(8) & 1.579(8)\\
  & (1,2) & 1.590(7) & 1.577(8)\\
4 & (0,4) & 1.637(10)  & 1.619(10)\\
  & (1,3) & 1.633(9) & 1.624(8)\\
  & (2,2) & 1.632(9) & 1.614(7)\\
5 & (0,5) & 1.694(10)  & 1.671(6)\\
  & (1,4) & 1.689(10)  & 1.667(5)\\
  & (2,3) & 1.688(9) & 1.667(3)\\
6 & (0,6) & 1.724(15)  & 1.689(11)\\
  & (2,4) & 1.712(16)  & 1.678(16)\\
  & (3,3) & 1.712(16)  & 1.678(17)\\
7 & (0,7) & 1.782(11)  & 1.756(10)\\
  & (3,4) & 1.766(14)  & 1.741(23)\\
8 & (0,8) & 1.843(8) & 1.846(19)\\
  & (4,4) & 1.834(4) & 1.835(10)\\
\hline\hline
\end{tabular}
\end{center}
\end{table}

\begin{table}[b]
\caption{\label{tab7}
The best-fit values of $\sigma_{\rm eff}$, $A_{\rm eff}$, and $C_{\rm eff}$ in Eq.~(\ref{VQQq}).
The list also shows the used gauge configuration number $N_{\rm conf}$ and their $\chi ^2$ over the degree of freedom $N_{\rm dof}$.
}
\begin{ruledtabular}
\begin{tabular}{cccccc}
$\kappa$ & $N_{\rm conf}$ & $\sigma_{\rm eff}$ & $A_{\rm eff}$ & $C_{\rm eff}$ & $\chi ^2/N_{\rm dof}$\\
\hline
0.1200 & 1000 & 0.045(2) & 0.12(2) & 1.49(2) & 1.31\\
0.1300 & 300  & 0.038(4) & 0.13(2) & 1.23(3) & 1.18\\
0.1340 & 300  & 0.037(4) & 0.13(2) & 1.12(2) & 1.11\\
0.1380 & 1000 & 0.037(2) & 0.13(1) & 0.97(1) & 1.16\\
\end{tabular}
\end{ruledtabular}
\end{table}

\subsection{Effective string tension}
\begin{figure}[t]
\begin{center}
\includegraphics[scale=0.4]{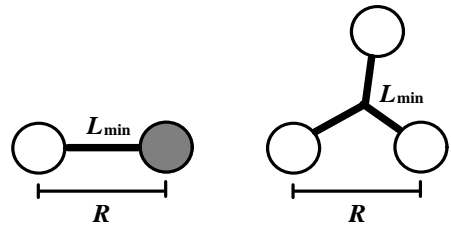}
\caption{\label{fig5}
The schematic figure of the flux-tube length $L_{\rm min}$ and the inter-two-quark distance $R$.
These are equal, i.e., $L_{\rm min} =R$, in the $Q\bar Q$ system (left), and these are not equal, i.e., $L_{\rm min} \neq R$, in the $3Q$ or $QQq$ system (right).
}
\end{center}
\end{figure}

In the ground state inter-quark potentials in hadrons, the confinement potential is the linear function of the flux-tube length $L_{\rm min}$, and the flux tube forms the shape minimally connecting the quarks.
As depicted in Fig.~\ref{fig5}, in the ground state of $Q\bar Q$ systems, the flux-tube length $L_{\rm min}$ equals to the distance $R$ between the quark and the antiquark.
The confinement part of the $Q\bar Q$ potential can be written as a linear function of the inter-quark distance $R$, and the string tension $\sigma _{Q\bar Q}$ is its proportionality coefficient.

In contrast, in $3Q$ systems or multi-quark systems, the flux tube length $L_{\rm min}$ and the inter-quark distances do not coincide, and its relation is determined by nontrivial dynamics of QCD.
Thus, the confinement potential is a linear function of $L_{\rm min}$ but a complicated function of the inter-quark distances.
In addition, the $QQq$ potential effectively includes the heavy-light Coulomb potential and the light-quark kinetic energy.
Therefore, the $R$-dependence of the $QQq$ potential itself, for example that the $QQq$ confinement potential is linear with $R$, is a nontrivial result.

The lattice QCD results suggest that the $QQq$ confinement potential is written as the familiar linear potential form, but the effective string tension $\sigma_{\rm eff}$ is smaller than the static $3Q$ string tension $\sigma_{3Q}$.
The string tension is the proportionality coefficient of the flux-tube length in confinement potentials, and characterizes the confining force by the flux tube.
The effective string tension is the proportionality coefficient of the inter-two-quark distance, and characterizes the confining force between two quarks in hadrons.
The deviation between the string tension and the effective string tension is considered to originate from such a difference, i.e., the geometrical difference between the flux-tube length and the inter-quark distance.

As mentioned before, the functional form of the $QQq$ potential is generally nontrivial.
Let us consider the more detail about its functional form.
In the large $R$ limit, the light-quark spreading vertical to the inter-heavy-quark direction is negligible compared with $R$, and the flux-tube length is approximately equal to $R$.
This intuitive consideration suggests that $\sigma_{\rm eff}$ would approach to $\sigma_{3Q}$ at much larger $R$, namely the effective string tension is some function of $R$, not a constant.
This behavior is rather natural, since the $QQq$ confinement potential is not necessarily a linear function of $R$ under the nontrivial relation between the flux tube length and the inter-quark distance.
If the $QQq$ confinement potential $V_{QQq}^{\rm conf}(R)$ is a general function of $R$, the effective string tension is defined as its derivative,
\begin{eqnarray}
\sigma_{\rm eff}(R) \equiv \frac{\partial V_{QQq}^{\rm conf}(R)}{\partial R},
\end{eqnarray}
and can depend on $R$.
In the potential model study of Ref.~\cite{Ya08}, by calculating up to $R=2.4$ fm, it is confirmed that the effective string tension slightly depends on $R$.
The same behavior will be confirmed also in lattice QCD with a larger-volume calculation.

Next we consider the light-quark-mass dependence of the $QQq$ potential.
When the light-quark mass is larger, the spatial extension is more compact and the flux-tube length is closer to the inter-heavy-quark distance.
In the infinite mass limit, the $QQq$ system corresponds to the static $3Q$ system, and the effective string tension equals to the string tension.
Then the effective string tension is an increasing function of the light-quark mass, and approaches asymptotically to $\sigma _{3Q}$ in the infinite mass limit.
We can confirm these behaviors in Table \ref{tab7}.
 
In Ref.~\cite{Ya08}, the same $QQq$ potential is investigated in a non-relativistic potential model, or a quark model.
This potential model reproduces the present lattice QCD result under the same condition of the quark mass and the range of $R$.
The potential model can calculate the light-quark wave function and the expectation value of the $QQq$ flux-tube length.
It enables us to understand the reduction mechanism of the effective string tension.
By investigating the relation between the flux-tube length and the inter-quark distance $R$ quantitatively, we confirm that a geometrical difference between these is essential for the reduction of the effective string tension, as conjectured above. 

We have found that, in $QQq$ systems, the effect of the finite-mass valence quark reduces the effective string tension between the two heavy quarks from the string tension of static $3Q$ systems. 
This reduction originates with the fact that the inter-quark distance differs from the flux-tube length in $3Q$ systems.
This is a simple and general property.
Our calculation is performed with $QQq$ systems for simplicity, however, this simple argument would also hold for ordinary baryons, which include three finite-mass quarks.
Although the finite-mass correction is more complicated, the effective string tension can be reduced in ordinary baryons, such as a nucleon.
Furthermore, this can be also applied to the multi-quark system including light quarks \cite{Gr96}.
In multi-quark systems, the inter-two-quark potential receives the more complicated effects of other valence quarks, and the effective string tension between the two quarks would be changed from the string tension.

\section{SUMMARY}
In summary, we have studied the $QQq$ potential in SU(3) lattice QCD, and investigated the role of the finite-mass valence quark in the inter-quark potential.
For the error reduction, we have adopted the Coulomb gauge fixing and the wall-to-wall quark propagator.
From the $Q\bar Q$ and $3Q$ potentials with the Coulomb gauge, the Coulomb gauge calculation approximately gives the physical potential in the whole region.

We have found that the $QQq$ potential is well described with a Coulomb plus linear potential, at least in the region of $R \le 0.8$ fm.
The Coulomb coefficient $A_{\rm eff}$ is almost the same as the $3Q$ case $A_{3Q}$, but interestingly, the effective string tension $\sigma _{\rm eff}$ is 10-20\% reduced from the string tension $\sigma _{3Q}$ in the static $3Q$ case.
The light-quark mass dependence of the potential is also investigated in the range of 0.5 GeV $\le M_q \le 1.5$ GeV.

The effective string tension is the confining force between two heavy quarks in $QQq$ systems.
The reduction of the effective string tension means that the inter-two-quark confining force appears to be weakened by the motional effect of the other finite-mass valence quark.
It originates from the difference between the flux-tube length and the inter-quark distance, and reflects the characteristic flux-tube structure of baryons.

This reduction of the inter-two-quark confinement force is conjectured to be a general property not only for $QQq$ systems but also for ordinary baryons.
Also in multi-quark hadrons, we can expect similar or more complicated effects on the inter-two-quark potential by the finite-mass valence quark.
The quark confinement is a fundamental property for hadrons, and its change would be important for broad fields relating quark-hadron physics.

\section*{ACKNOWLEDGEMENTS}
We thank Dr.~T.~T.~Takahashi, Dr.~T.~Umeda and Prof.~S.~J.~Brodsky for useful comments and discussions.
A.~Y.~and H.~S.~are supported by a Grant-in-Aid for Scientific Research [(C) No.~20$\cdot$363 and (C) No.~19540287] in Japan.
H.~I.~is supported by Yukawa International Program for Quark-Hadron
Sciences (YIPQS).
The lattice QCD calculations are done on NEC SX-8R at Osaka University.

\appendix

\section{Other trials on statistical improvement}
We have tried several statistical improvement techniques for calculating the $QQq$ potential.
These techniques are useful for error reduction to some extent, but not enough to calculate the $QQq$ potential in gauge-invariant manner.
The techniques are not used for the final result of the Coulomb gauge calculation.
We introduce them briefly in this Appendix.

\begin{figure}[t]
\begin{center}
\includegraphics[scale=0.4]{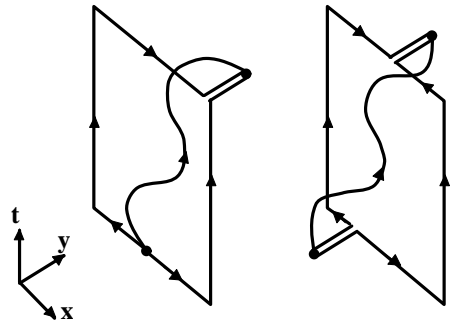}
\caption{\label{fig6}
Examples of the junction choices of the $QQq$ Wilson loop.
The wavy line represents the light-quark propagator and the straight line the heavy-quark trajectory.
}
\end{center}
\end{figure}

\subsection{Multi-hit procedure}
We have applied the multi-hit procedure \cite{Pa83,Ba95}.
The temporal link variables are replaced with the mean-field value of the neighboring link variables.
It is realized by replacing the temporal link variable $U_4(n)$ in the staples with $\tilde{U}_4(n)(\in $ SU(3)) which maximizes
\begin{eqnarray}
&\sum_j{\rm ReTr} \Bigl[ \tilde{U}_4^{\dagger}(n) \bigl\{ U_j(n)U_4(n+\hat{j})U_j^\dagger (n+\hat{4})\nonumber\\
&+U_j^\dagger (n-\hat{j})U_4(n-\hat{j})U_j(n+\hat{4}-\hat{j}) \bigr\} \Bigr]. 
\end{eqnarray}

\subsection{Average of the junction points}
In Fig.~\ref{fig1}, the positions $(R_{\rm I}, R_{\rm II})$ of the junction points are the same at the source and sink of the $QQq$ Wilson loop.
We can take $(R_{\rm I}, R_{\rm II})$ at the source and sink independently, and average all the combinations of $(R_{\rm I}, R_{\rm II})$ with fixed $R$.
This improvement increases the statistics by $(R+1)^2$ times, and effective in large $R$, where the statistical error is severe.

\subsection{More average of the junction points}
The positions of the junction points can be taken more arbitrarily.
For example, as depicted in Fig.~\ref{fig6}, we have perpendicularly bended the path of the spatial links in the staples, and averaged such contributions with fixed $R$.
In addition, we would be able to take more arbitrary shapes of staples, or off-axis $QQq$ Wilson loops.

\section{Gauge dependence}
We show here the results of the $QQq$ potential without gauge fixing or with another gauge.

\begin{figure}[t]
\begin{center}
\includegraphics[scale=1.1]{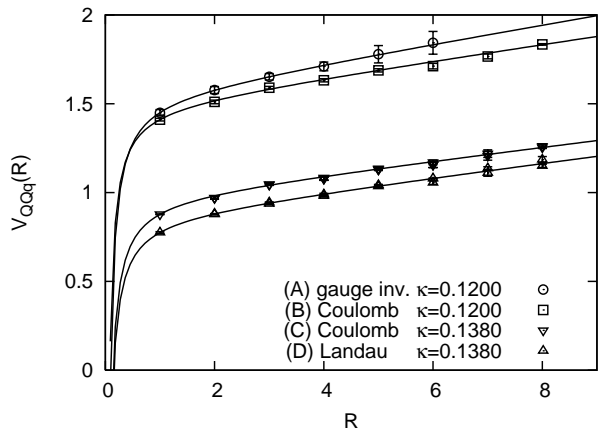}
\caption{\label{fig7}
The $QQq$ potential $V_{QQq}$ of the different gauge choices:
(A) gauge invariant with $\kappa=0.1200$ and $N_{\rm conf}=4000$,
(B) Coulomb gauge with $\kappa=0.1200$ and $N_{\rm conf}=1000$,
(C) Coulomb gauge with $\kappa=0.1380$ and $N_{\rm conf}=1000$,
(D) Landau gauge with $\kappa=0.1380$ and $N_{\rm conf}=200$.
The solid curves are the best-fit functions of Eq.~(\ref{VQQq}).
}
\end{center}
\end{figure}

\begin{table}[t]
\caption{\label{tab8}
The results of the different gauge choice.
The list shows the configuration number $N_{\rm conf}$, and the best-fit values of $\sigma_{\rm eff}$ and $A_{\rm eff}$ in Eq.~(\ref{VQQq}).
A statistical improvement is adopted  at the gauge-invariant calculation.
}
\begin{ruledtabular}
\begin{tabular}{ccccc}
gauge & $\kappa$ & $N_{\rm conf}$ & $\sigma_{\rm eff}$ & $A_{\rm eff}$\\
\hline
gauge inv. & 0.1200 & 4000 & 0.051(3) & 0.15(1) \\
Coulomb    & 0.1200 & 1000 & 0.045(2) & 0.12(2) \\
Coulomb    & 0.1380 & 1000 & 0.037(2) & 0.13(1) \\
Landau     & 0.1380 & 200  & 0.039(2) & 0.13(2) \\
\end{tabular}
\end{ruledtabular}
\end{table}

We calculate the gauge-invariant $QQq$ potential in the large-quark-mass case $\kappa=0.1200$ with the large statistics $N_{\rm conf}=4000$.
In the large-quark-mass case, compared to the small-quark-mass case, the statistical error is relatively small and the computing time for the quark propagator is fairly short.
For further statistical improvement, the junction-average procedure explained in Appendix A2 is adopted for the gauge-invariant calculation.
Nonetheless, the statistical error is still large, and its data can be taken only in the region of $R\le 0.6$ fm.
The result is shown in Fig.~\ref{fig7} and Table \ref{tab8}.
Note that the listed error in Table \ref{tab8} is only the statistical error, and that the gauge-invariant calculation involves the systematic error from the fit-range dependence.
The gauge-invariant result seems to be close to the Coulomb gauge result, but the precise comparison is difficult by the large statistical and systematic error.

In the lighter quark case, the statistical error is severely large and the potential cannot be extracted at all.
Instead, we show the Landau gauge result at $\kappa=0.1380$ in Fig.~\ref{fig7} and Table \ref{tab8}.
The result with the Landau gauge is roughly coincident to that with the Coulomb gauge, except the irrelevant constant shift.


\begin{thebibliography}{99}
\bibitem{Cr81} M.~Creutz, {\it Quarks, Gluons and Lattices} (Cambridge University Press, Cambridge, England, 1983).
\bibitem{Ro92} H.~J.~Rothe, {\it Lattice Gauge Theories} (World Scientific, Singapore, 1992).
\bibitem{Na74} Y.~Nambu, Phys. Rev. D {\bf 10}, 4262 (1974).
\bibitem{Ei78} E.~Eichten, K.~Gottfried, T.~Kinoshita, K.~D.~Lane, T.-M.~Yan, Phys. Rev. D {\bf 17}, 3090 (1978).
\bibitem{Cr7980} M.~Creutz, Phys. Rev. Lett. {\bf 43}, 553 (1979); Phys. Rev. D {\bf 21}, 2308 (1980).
\bibitem{Ba92} G.~S.~Bali and K.~Schilling, Phys. Rev. D {\bf 46}, 2636 (1992).
\bibitem{Ta0102} T.~T.~Takahashi, H.~Matsufuru, Y.~Nemoto, and H.~Suganuma, Phys. Rev. Lett. {\bf 86}, 18 (2001); T.~T.~Takahashi, H.~Suganuma, Y.~Nemoto, and H.~Matsufuru, Phys. Rev. D {\bf 65}, 114509 (2002).
\bibitem{Ic03} H.~Ichie, V.~Bornyakov, T.~Streuer, and G.~Schierholz, Nucl. Phys. {\bf A721}, 899 (2003); V.~G.~Bornyakov, H.~Ichie, Y.~Mori, D.~Pleiter, M.~I.~Polikarpov, G.~Schierholz, T.~Streuer, H.~Stuben, and T.~Suzuki, Phys. Rev. D {\bf 70}, 054506 (2004).
\bibitem{Br95} N. Brambilla, G. M. Prosperi, and A. Vairo, Phys. Lett. B {\bf 362}, 113 (1995).
\bibitem{Co04} J.~M.~Cornwall, Phys. Rev. D {\bf 69}, 065013 (2004).
\bibitem{Ok05} F.~Okiharu, H.~Suganuma, and T.~T.~Takahashi, Phys. Rev. Lett. {\bf 94}, 192001 (2005); Phys. Rev. D {\bf 72}, 014505 (2005).
\bibitem{Ba05} G.~S.~Bali, H.~Neff, T.~Dussel, T.~Lippert, and K.~Schilling, Phys. Rev. D {\bf 71}, 114513 (2005).
\bibitem{Ru75} A.~De Rujula, H.~Georgi, and S.~Glashow, Phys. Rev. D {\bf 12}, 147 (1975).
\bibitem{YaL} A.~Yamamoto, H.~Suganuma, and H.~Iida, Phys. Lett. B {\bf 664}, 129 (2008).
\bibitem{Ma02} M.~Mattson {\it et al.} (SELEX Collaboration), Phys. Rev. Lett. {\bf 89}, 112001 (2002).
\bibitem{Oc05} A.~Ocherashvili {\it et al.} (SELEX Collaboration), Phys. Lett. B {\bf 628}, 18 (2005).
\bibitem{Au06} B.~Aubert {\it et al.} (BABAR Collaboration), Phys. Rev. D {\bf 74}, 011103 (2006).
\bibitem{Le0102} R.~Lewis, N.~Mathur, and R.~M.~Woloshyn, Phys. Rev. D {\bf 64}, 094509 (2001);  N.~Mathur, R.~Lewis, and R.~M.~Woloshyn, Phys. Rev. D {\bf 66}, 014502 (2002).
\bibitem{Vi04} J.~Vijande, H.~Garcilazo, A.~Valcarce, and F.~Fernandez, Phys. Rev. D {\bf 70}, 054022 (2004).
\bibitem{Br05} N.~Brambilla, A.~Vairo, and T.~Rosch, Phys. Rev. D {\bf 72}, 034021 (2005).
\bibitem{Bo92} S.~P.~Booth {\it et al.} (UKQCD Collaboration), Phys. Lett. B {\bf 275}, 424 (1992).
\bibitem{El97} A.~X.~El-Khadra, A.~S.~Kronfeld, and P.~B.~Mackenzie, Phys. Rev. D {\bf 55}, 3933 (1997);
G.~P.~Lepage and P.~B.~Mackenzie, Phys. Rev. D {\bf 48}, 2250 (1993).
\bibitem{Gr03} J.~Greensite and S.~Olejnik, Phys. Rev. D {\bf 67}, 094503 (2003).
\bibitem{Ya08} A.~Yamamoto and H.~Suganuma, Phys. Rev. D {\bf 77}, 014036 (2008).
\bibitem{Gr96} A.~M.~Green, J.~Lukkarinen, P.~Pennanen, and C.~Michael, Phys. Rev. D {\bf 53}, 261 (1996); P.~Pennanen, A.~M.~Green, and C.~Michael, Phys. Rev. D {\bf 59}, 014504 (1998).
\bibitem{Pa83} G.~Parisi, R.~Petronzio, and F.~Rapuano, Phys. Lett. B {\bf 128}, 418 (1983).
\bibitem{Ba95} G.~S.~Bali, C.~Schlichter, and K.~Schilling, Phys. Rev. D {\bf 51}, 5165 (1995).
\end{thebibliography}
\end{document}